\documentclass[preprint]{elsarticle}
\graphicspath{ {./figures/} }
\usepackage{hyperref}
\usepackage{float}
\usepackage{verbatim} 
\usepackage{apalike}
\restylefloat{figure}
\floatstyle{plaintop} 
\restylefloat{table}
\usepackage{amsmath}
\usepackage{upgreek}
\usepackage{blindtext}
\usepackage{graphicx}
\usepackage{siunitx}
\usepackage{float}
\journal{Expert Systems with Applications}

\biboptions{authoryear}

\begin{document}
\begin{frontmatter}

\title{Inverse-Designed Metasurfaces for Wavefront Restoration in Under-Display Camera Systems}

\author[label1,label2]{Jaegang Jo}
\author[label3,label2]{Myunghoo Lee}
\author[label1,label2]{Seunghyun Lee}
\author[label1]{Munseong Bae}
\author[label4]{Chanik Kang}
\author[label1,label4]{Haejun Chung \corref{cor1}}
\ead{haejun@hanyang.ac.kr}

\cortext[cor1]{Corresponding author.}
\address[label1]{Department of Electronic Engineering, Hanyang University, Seoul, 04763, South Korea}
\address[label3]{Department of Electrical and Computer Engineering, University of Washington, Seattle, WA, 98195, USA }
\address[label4]{Department of Artificial Intelligence, Hanyang University, Seoul, 04763, South Korea}
\address[label2]{These authors contributed equally to this work.}

\begin{abstract}
Under-display camera (UDC) systems enable full-screen displays in smartphones by embedding the camera beneath the display panel, eliminating the need for notches or punch holes. However, the periodic pixel structures of display panels introduce significant optical diffraction effects, leading to imaging artifacts and degraded visual quality. Conventional approaches to mitigate these distortions, such as deep learning-based image reconstruction, are often computationally expensive and unsuitable for real-time applications in consumer electronics. This work introduces an inverse-designed metasurface for wavefront restoration, addressing diffraction-induced distortions without relying on external software processing. The proposed metasurface effectively suppresses higher-order diffraction modes caused by the metallic pixel structures, restores the optical wavefront, and enhances imaging quality across multiple wavelengths. By eliminating the need for software-based post-processing, our approach establishes a scalable, real-time optical solution for diffraction management in UDC systems. This advancement paves the way to achieve software-free real-time image restoration frameworks for many industrial applications.
\end{abstract}


\end{frontmatter}

\section{Introduction}
\label{introduction}
In next-generation display technologies, including those used in AR/VR systems, smartphones, and tablets, there is an increasing demand for display-integrated imaging systems. These systems aim to deliver seamless visual experiences while preserving high-quality imaging capabilities. However, conventional displays are typically composed of metallic pixel arrays, which can reflect or diffract incident light, resulting in significant image distortion at the camera module. To address this issue, under-display camera (UDC) technology has emerged as a promising solution, integrating camera modules beneath partially transparent OLED or MicroLED displays to achieve full-screen designs with minimal visual disruption \citep{Kwon16,Qin16}. By utilizing partially transparent displays, UDCs maintain high screen-to-body ratios and provide an uninterrupted viewing experience. These advantages make UDC technology a compelling candidate for next-generation electronic devices\citep{Lim2020}. Nevertheless, despite its potential, UDC technology continues to face considerable optical challenges that must be addressed to achieve optimal imaging performance.

One of the most significant challenges in UDC systems is the control of the optical diffraction caused by the periodic apertures of the metallic pixels on the display panels\citep{Qin17,tang20}. These optical diffraction patterns generate unwanted artifacts on the image sensor, degrading image quality, and diminishing overall system performance. This issue is especially problematic in applications that require high-resolution imaging, such as biometric authentication and augmented reality systems, where image clarity is crucial for functionality. Effectively mitigating unwanted optical diffraction while maintaining the quality of the screen output in the display remains a fundamental technical barrier.

Various solutions have been explored to address these challenges. Deep learning-based image restoration algorithms have shown promise in compensating for degraded images, including diffraction-induced image distortions\citep{Zhou_2021_CVPR,Kwon_2021_CVPR,Song_2023_ICCV,Luo22}. However, these methods are often computationally intensive, making them unsuitable for real-time applications. One of the hardware-based solutions is to place the display pixels in slightly random positions; thereby, the optical diffractions do not form a strong concentration to higher order modes at the image sensor\citep{yang2021designing}. However, this solution degrades the overall imaging quality due to the random optical diffractions. Also, it compromises the uniformity of the display pixels, leading to uncomfortable screen output. 

In this work, we introduce inverse-designed metasurfaces as a wavefront restoration technique to effectively eliminate unwanted optical diffraction in UDC systems. Metasurfaces offer unparalleled control over light at sub-wavelength scales, enabling precise manipulation of phase, amplitude, and polarization\citep{Huang18,Jiang19,kang2024large}. Their ultrathin and planar form factor minimizes the size and weight of optical components, making them well-suited for compact electronic devices. Additionally, metasurfaces can be engineered for broadband operation across multiple wavelengths, facilitating their integration into a wide range of applications, including lenses~\citep{Liu20}, vortex beam generators~\citep{Yue16}, holograms~\citep{wan17}, beam splitters~\citep{Khorasaninejad15}, and beam steerers~\citep{Naqvi19}. Their compatibility with nanofabrication techniques further enhances scalability, supporting practical deployment in consumer electronics.

To achieve highly specialized metasurface designs for wavefront restoration, we employ the adjoint method—an efficient large-scale inverse design technique~\citep{hughes2018,wang2020,li2022empowering,kang2024adjoint}. This method computes the derivatives of the figure of merit with respect to design parameters, leveraging the Born approximation and Lorentz reciprocity. By requiring only two full-wave simulations (forward and adjoint), the adjoint method enables the computation of billions of derivatives, significantly reducing the computational burden associated with optimization. This approach has successfully addressed various optical and photonic design challenges, including the development of metalenses~\citep{chung2020}, holograms~\citep{yin2024multi}, beam demultiplexers~\citep{piggott2015inverse}, and applications in quantum computing~\citep{goel2024inverse}.

We develop an adjoint optimization framework for controlling optical diffraction power in periodic structures and apply it to the design of a metasurface that converts randomized incident waves into a zeroth-order plane wave. We then address unwanted optical diffraction in UDC systems, where the periodic arrangement of display pixels is modeled as periodic silver slabs, and the input image is assumed to be a zeroth-order plane wave. The optimized multi-layer metasurface effectively mitigates diffraction patterns induced by the metallic slabs, thereby enhancing imaging quality in UDC systems without the need for complex image restoration models. This breakthrough enables real-time image restoration without software dependencies, opening new possibilities for various industrial applications.

The key contributions of this work are summarized below:
\begin{enumerate}
\item We propose an adjoint-based inverse design framework to develop multi-layer metasurfaces capable of precise wavefront control and effective suppression of higher-order optical diffraction.
\item We propose a software-free, real-time solution to diffraction-induced image degradation in UDC systems.
\item We demonstrate the highest reported zeroth-order mode purity (near-perfect restoration) for arbitrary incidence waveform.
\item The proposed system is quantitatively validated through point spread function (PSF) and modulation transfer function (MTF) analyses, confirming substantial improvements in spatial frequency preservation and imaging quality.
\end{enumerate}

This study highlights the potential of intelligent optical system design driven by optimization for consumer electronics. It offers a practical and scalable solution for real-time wavefront restoration in embedded vision systems. Furthermore, the proposed methodology is broadly applicable to other domains requiring compact optical computing, diffraction management, and advanced photonic system design.

The remainder of this article is organized as follows: Section 2 introduces the adjoint-based optimization framework for metasurface design and presents its performance in suppressing higher-order diffraction. Section 3 demonstrates the integration of the optimized metasurface into UDC systems, along with a comprehensive evaluation of wavefront restoration and imaging quality under realistic conditions. Finally, Section 4 concludes the paper with a summary of key findings and potential directions for future research.

\section{Metasurface designs for wavefront correction}
Subwavelength-scale photonic structures arranged in periodic configurations inherently redistribute incident power among multiple diffraction orders. Maintaining wavefront during propagation is essential to preserving spatial information in many imaging applications. Conventional wavefront correction relies on active optical elements such as deformable mirrors~\citep{cahoy2014wavefront}, spatial light modulators~\citep{spangenberg2014white}, and liquid crystal~\citep{serati2005advances, huang2018wavefront, cao2008correction} or MEMS-based devices~\cite{poyneer2008laboratory, stewart2007open}, which dynamically modify the phase of incident light to allow programmable beam shapings, holography, optical tweezers, and wavefront sensings. However, these approaches involve bulky active components, making them impractical for integration into emerging display-camera systems.
In contrast, metasurfaces leverage subwavelength-scale structures to precisely control phase, amplitude, and polarization, enabling efficient suppression of undesired diffraction modes and directing energy into the desired diffraction orders. Their ultrathin form factor, comparable to the operating wavelength, makes them well-suited for compact imaging systems such as smartphones, AR/VR devices, and unmanned aerial vehicles. 
In this section, we develop an adjoint-based topology optimization framework to design a metasurface that precisely manipulates complex higher-order diffraction components, converting them into a zeroth-order plane wave. Finally, numerical results demonstrate that the optimized metasurface efficiently converts higher-order diffraction modes into the desired zeroth-order transmitted wavefront.

\subsection{Power computation in transmitted diffraction orders}
To fully harness the potential of metasurface-based wavefront shaping, it is crucial to quantify how incident power is distributed among different transmitted diffraction orders. We consider a two-dimensional periodic metasurface lying in the \(xy\)-plane, illuminated by a monochromatic plane wave from above. Due to the periodicity of the structure, the transmitted field in the substrate region must satisfy the Bloch-periodic boundary conditions\citep{chutinan2000waveguides} and it can be decomposed into a series of optical diffraction orders, each characterized by a distinct wave vector. Accordingly, the transmitted electric field, \(\mathbf{E}_\text{trans}\), can be expressed as a superposition of plane waves:  
\begin{equation}
    \mathbf{E}_\text{trans}(\mathbf{r}) = \sum_{m\sigma} C_{m\sigma} \hat{\boldsymbol{\varepsilon}}_{m\sigma} \exp\left({i \mathbf{k}_m \cdot \mathbf{r}}\right),
\end{equation}
where \( C_{m\sigma} \) represents the complex amplitude of the $m _\text{th}$ order plane wave with polarization state \( \sigma \), \( \hat{\boldsymbol{\varepsilon}}_{m\sigma} \) is the polarization unit vector, and \( \mathbf{k}_m \) is the corresponding wave vector for the $m _\text{th}$ order plane wave. The wave vector of the transmitted optical diffraction orders satisfies the Bloch periodicity boundary condition:
\begin{equation}
    k_{mx} = k_x + m \frac{2\pi}{\Lambda_x},\quad k_{ny} = k_y + n \frac{2\pi}{\Lambda_y},
\end{equation}
where \( \Lambda_x \) and \( \Lambda_y \) define the periodicity of the unit cell along the \( x \)-and \(y \)-directions, respectively. The \(z \) component of the transmitted wavevector is given by:
\begin{equation} 
k_{mz} = \sqrt{\epsilon_r \frac{\omega^2}{c^2} - |\mathbf{k}_m \cdot \boldsymbol{\rho}|^2},
\end{equation}
where \(\epsilon_r\) is the background permittivity and \(\boldsymbol{\rho}\) is the projection operator onto the transverse plane, such that  \(|\mathbf{k}_m \cdot \boldsymbol{\rho}|^2= k_{mx}^2+k_{my}^2\). If \(k_{mz}\) becomes imaginary, the corresponding optical diffraction order is evanescent and does not contribute to the transmitted power in the far field. To quantify the contribution of each diffraction order, the transmitted electric field is projected onto the corresponding basis functions, allowing for the extraction of the amplitude of each plane wave component associated with a specific optical diffraction mode. 

\begin{equation} 
C_{m\sigma} = \frac{1}{A} \int_S \exp\left({-i\mathbf{k}_m \cdot \mathbf{r}}\right) \mathbf{E}_\text{trans}(\mathbf{r}) \cdot \hat{\boldsymbol{\varepsilon}}^*_{m\sigma} dA, 
\end{equation}

where $A$ represents the area of the integration region on the plane $S$ in the transmission region. The orthogonality of plane waves in periodic structures ensures that only modes with matching wavevectors contribute to \( C_{m\sigma} \), while all other components vanish, enabling a distinct separation of optical diffraction orders in the transmitted field.
By substituting the plane wave expansion of the transmitted field into the integral and applying the orthogonality condition, the expression for the transmitted power in each diffraction order simplifies to:
\begin{equation} 
\begin{aligned}
P_{m\sigma} &= \frac{1}{2} \operatorname{Re} \int_S \mathbf{E}_{{m\sigma}} \times \mathbf{H}_{m\sigma}^* \cdot \hat{z} \, dA \\
&= \frac{A}{2Z_t} |C_{m\sigma}|^2 \operatorname{Re}\left(\frac{k_{mz}}{|\mathbf{k}_m|}\right).
\end{aligned}
\end{equation}
where $Z_t$ is the wave impedance in the transmitted medium, which is given by $\sqrt{\mu/\epsilon}$. This formulation ensures that the power carried by each transmitted diffraction order is properly accounted for, providing an accurate assessment of how incident energy is redistributed among diffraction modes.

\subsection{Adjoint optimization for correcting diffraction orders} \label{sec:2.2}

\begin{figure}
    \centering
    \includegraphics[width=1.0\linewidth]{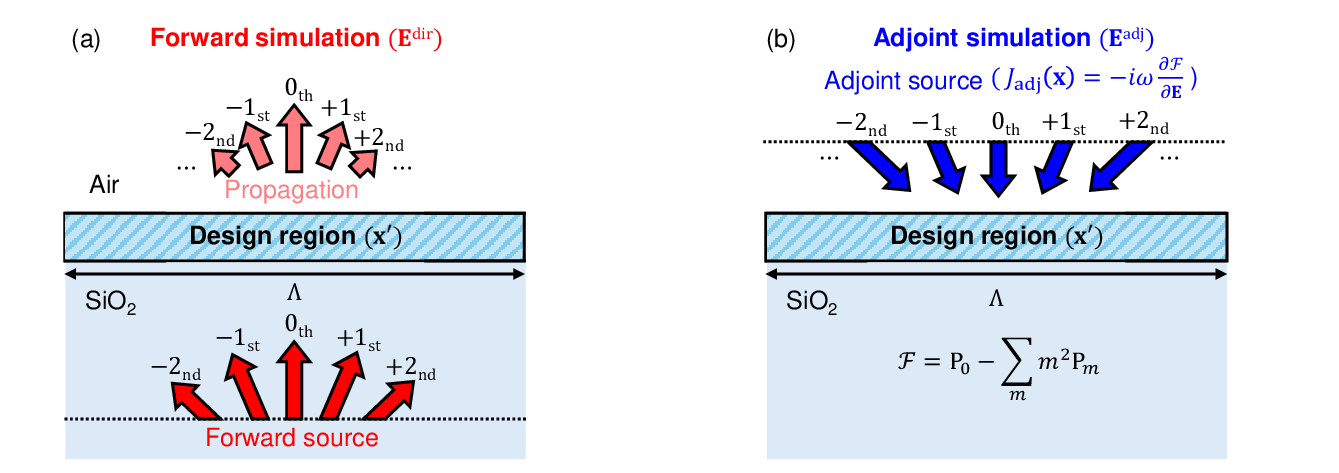}
    \caption{Schematic illustration of the forward and adjoint simulations for adjoint optimization of free-form metasurfaces. (a) Forward simulation: Incident plane waves interact with subwavelength nano-structures in the design region, redistributing the incident power into various diffracted plane wave modes. The resulting power distribution serves as the basis for the subsequent adjoint simulation. 
    (b) Adjoint simulation: Adjoint sources are introduced at the diffraction field monitor to compute the sensitivity of the FoM with respect to design parameters. The back-propagated field distribution aligns with the optimization objective, enabling efficient gradient computation for large-scale optimization.}
    \label{fig:fig1}
\end{figure}

Global optimization methods, such as genetic algorithms\citep{li2018broadband,liu2020genetic} and particle swarm optimization\citep{forestiere2010particle}, require a large number of simulations to explore the design space, making them computationally expensive and slow to converge. In contrast, semi-analytical methods, like rigorous coupled-wave analysis \citep{colburn2021inverse}, are typically limited to simple geometries and idealized conditions, restricting their applicability to more complex structures. Data-driven approaches, including machine learning-based optimization, can efficiently approximate optimal designs\citep{so2020deep,wang2022inverse}; however, they often demand vast amounts of high-quality training data, which makes them resource-intensive and challenging to apply in untrained conditions. In comparison, adjoint-based optimization has emerged as a highly effective alternative, requiring only two simulations, forward and adjoint, to compute the gradient of the figure of merit (FoM) with respect to the design parameters. 

Free-form metasurface designs~\citep{park2022free} is often considered a challenging problem due to the potentially vast number of design variables, which can exceed several billion across the entire metasurface. Additionally, a deep understanding of the interactions between the nanostructures is required. Furthermore, performance analysis typically involves evaluating the far-field distribution of the electromagnetic field. To address these challenges, we propose an adjoint optimization method in combination with a decomposition of the diffracted plane wave modes in a near-field regime. This approach potentially enhances both the computational efficiency and the performance of the metasurface devices compared to their conventional optimization counterparts.

In Fig.~\ref{fig:fig1}(a), the forward simulation involves incident plane waves with various combinations of diffracted plane waves, where the diffraction angle is defined as $\theta_m=\sin^{-1}\left(\frac{m\lambda}{\Lambda}\right)$. Here, $\Lambda$ and $\lambda$ denote the periodicity and operating wavelength, respectively. The incident wave interacts with the subwavelength nano-structures in the design region, redistributing the incident power into a new distribution. The resulting power distribution serves as the basis for the subsequent adjoint simulation, enabling gradient evaluation.

In Fig.~\ref{fig:fig1}(b), the adjoint simulation exploits the reciprocal nature of Maxwell's equations to introduce adjoint sources at the diffraction field monitor where the FoM is evaluated, ensuring that the back-propagated field distribution aligns with the optimization objective. These sources are placed to compute the sensitivity of the FoM with respect to design parameters. By achieving full-wave Maxwell’s solutions in this backward simulation, the adjoint method effectively determines the gradient information over the entire design regime required for the gradient-based optimization, enabling large-scale computational optimizations.

The primary objective of this optimization process is to manipulate optical diffraction orders to direct incident light predominantly into the zeroth-order diffraction mode while minimizing undesired transmission into higher-order modes. The FoM is mathematically defined as:
\begin{equation}
\mathcal{F} =  P_{0z} - \sum_{m \neq 0} m^2P_{mz} ,
\label{eq:FOM}
\end{equation}
where $P_{mz}$ represents the power in the $m{\text{th}}$ diffraction mode with \(z\)-polarization electric field, as discussed in the previous section. In Eq.~\ref{eq:FOM}, the penalty term $m^2$ ensures that higher-order optical diffraction modes are effectively minimized. This penalty term enables precise wavefront control by focusing the transmitted light into the desired optical diffraction order.

To optimize the metasurface based on this objective, the sensitivity of the FOM with respect to the design parameters must be computed. This sensitivity, determined by the interaction between the forward and the adjoint field responses, is quantified by the gradient $\frac{\partial\mathcal{F}}{\partial\epsilon_i}$, which measures the influence of permittivity $\epsilon_i$ at each pixel within the design region. The gradient is mathematically expressed as~\citep{tang2023time}:
\begin{equation}
\frac{\partial \mathcal{F}}{\partial \epsilon_i} = \frac{1}{\pi} \text{Re} \left[ \int_{\Delta \omega} \omega^2 \mathbf{E}^{\text{adj}} (\mathbf{x}_i, \omega) \cdot \mathbf{E}{^{\text{dir}}}^{*}(\mathbf{x}_i, \omega) \, d\omega \right],
\label{eq:gradient}
\end{equation} where $\epsilon_{i}$ and $\mathbf{x}_i$ represent the permittivity and the spatial position of the $i_\text{th}$ pixel in the design region, respectively. The term $\mathbf{E}^\text{adj}$ denotes the adjoint electric field response, derived from an adjoint source, $\mathbf{E}{^{\text{dir}}}^{*}$ describes the complex conjugate of the forward electric field. The parameters $\omega$ and $\Delta\omega$ correspond to the angular frequency of the electromagnetic wave and the frequency range considered in the optimization, respectively.

To compute this gradient, the adjoint method introduces a set of adjoint sources $\mathbf{J}_{\text{adj}}\left(\mathbf{x}\right)$ at the diffraction field monitor where the FoM is evaluated. These sources ensure that the back-propagated field distributions align with the optimization objective. The adjoint source is mathematically defined as: 
\begin{equation} 
\mathbf{J}_{\text{adj}}\left(\mathbf{x}\right) = -i\omega\frac{\partial\mathcal{F}}{\partial\mathbf{E}}=-i\omega \left[ \mathbf{E}_0^*\left(\mathbf{x}\right) - \sum_{m \neq 0} \cos \theta_m \mathbf{E}_m^*(\mathbf{x}) \right],
\label{eq:adjoint source}
\end{equation}
where $\mathbf{E}_m^*\left(\mathbf{x}\right)$ represents the complex conjugate of the forward electric field corresponding to the $m_\text{th}$ optical diffraction mode at position $\mathbf{x}$, and $\theta_m$ denotes the propagation angle of the $m_\text{th}$ optical diffraction mode relative to the optical axis. 
By iteratively updating the permittivity distribution using a gradient descent algorithm, the metasurface design is progressively optimized to maximize the FoM. The permittivity update follows the rule: 
\begin{equation} 
\epsilon_i^{(n+1)} = \epsilon_i^{(n)} - \eta \frac{\partial\mathcal{F}}{\partial \epsilon_i}, 
\label{eq:epsilon_update} 
\end{equation} 
where $\eta$ is the step size and $n$ denotes the iteration number. Initially, the permittivity values are continuously varied, allowing the optimization process to explore a broad range of design space. As optimization progresses, the permittivity distribution gradually converges to the discrete values corresponding to the two materials, $\epsilon_{\text{SiO}_2}$ and $\epsilon_{\mathrm{Si_3N_4}}$, which represent the physical constraints of the design. This continuous-to-discrete transition ensures that the final metasurface design not only achieves optimal optical diffraction control but also remains manufacturable within practical nanofabrication limits.

\begin{figure}[H]
    \centering
    \includegraphics[width=1.0\linewidth]{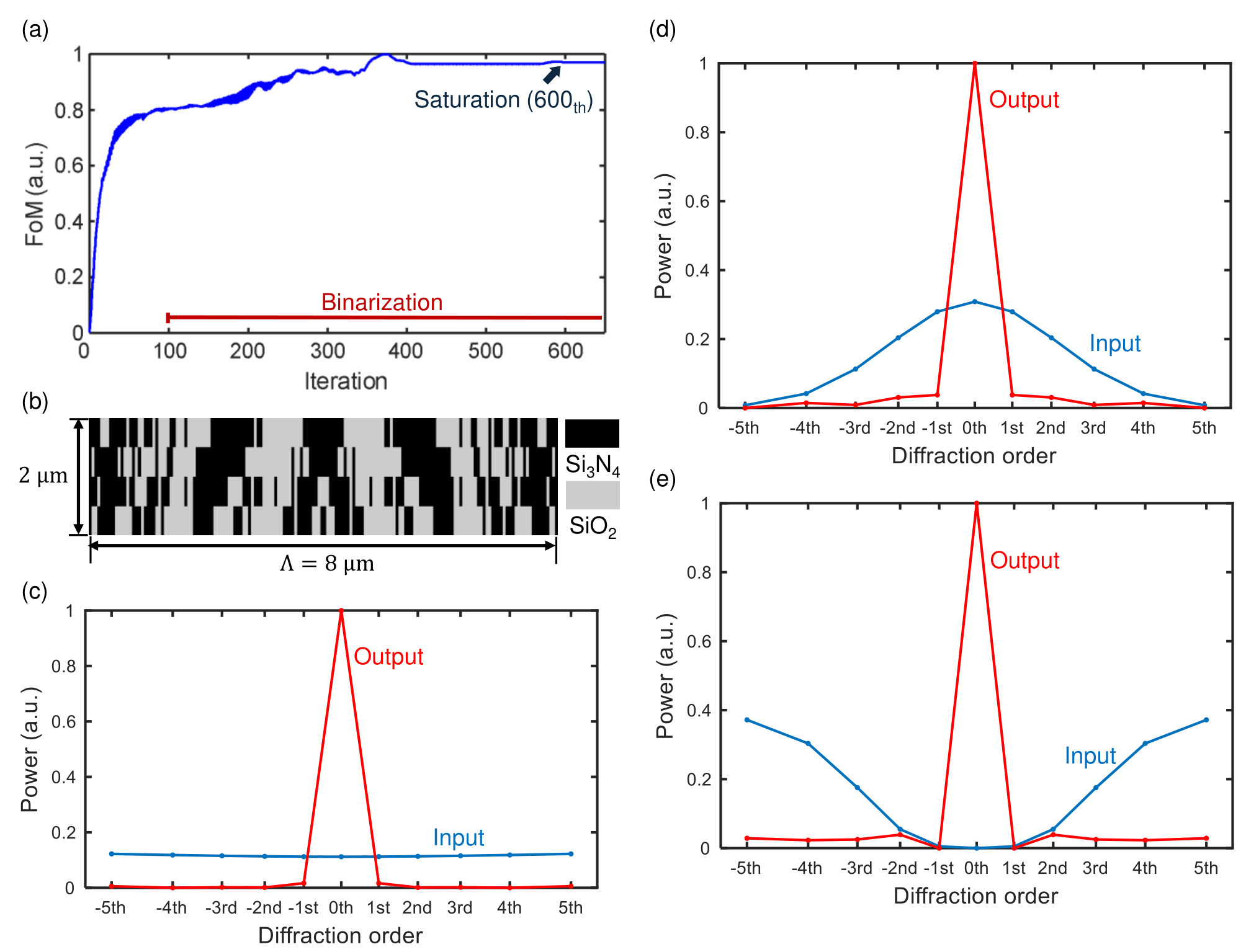}
    \caption{Optimization result and performance evaluation of the adjoint-based metasurface design. (a) Evolution of the figure of merit (FoM) over 600 adjoint optimization iterations, demonstrating convergence and binarization. (b) The optimized four-layer metasurface structure, composed of \(\mathrm{Si_3N}_4\) and \(\mathrm{SiO_2}\), with a unit length of 8~\unit{\micro\metre} and a thickness of 500~\unit{\nano\metre}. (c) Transmitted diffraction power distribution under a uniform input power distribution across diffraction orders (blue), showing a strong concentration of power in the zeroth-order mode after optimization (red). (d) Transmitted diffraction power distribution under a Gaussian-distributed input power across diffraction orders, demonstrating that the optimized metasurface successfully directs power into the zeroth-order mode despite varying input conditions. (e) Transmitted diffraction power distribution under an inverted Gaussian distribution of input diffraction plane waves, showing the robustness of metasurface for redistributing higher-order diffraction into the zeroth-order mode.}
    \label{fig:fig2}
\end{figure}

\subsection{Optimization results}
This section presents the results of the adjoint optimization applied to metasurface design, focusing on optical diffraction control, particularly when multiple diffraction orders are used as incidence waves. Figure~\ref{fig:fig2} illustrates the optimization process, the final metasurface design, and its resulting performance in terms of optical diffraction control and power distribution. Figure~\ref{fig:fig2}(a) illustrates the evolution of the FoM over 600 adjoint optimization iterations. Each iteration consists of two simulations (forward and adjoint), resulting in a total of 1300 simulations for the entire optimization process. The computational cost could be further reduced by optimizing the step size for parameter updates and incorporating non-linear optimization techniques. We strongly believe that other gradient-free optimizations would struggle to achieve the same high-quality optimal structure presented in this work, as they are often limited in their applicability to high-dimensional problems. Initially, the FOM increases sharply, reflecting rapid convergence toward the design objective. As optimization progresses, the rate of improvement gradually decreases, and the FOM asymptotically approaches a saturation point. As discussed in the previous section, the optimization process includes a binarization constraint to guarantee manufacturability while preserving optical performance. During this binarization, the permittivity distribution shifts from continuous values to discrete material domains, progressively evolving into either \(\mathrm{Si_3N_4}\) or \(\mathrm{SiO_2}\). The optimized structure, shown in Fig.~\ref{fig:fig2}(b), consists of four layers of the \(\mathrm{Si_3N_4}\) and \(\mathrm{SiO_2}\) structures. Each layer has 500~\unit{\nano\metre} thickness, while the minimum width of the nanostructure is 50~\unit{\nano\metre}, resulting in a maximum aspect ratio of the nanostructure to 10.

We first apply a uniform power distribution across all possible diffraction orders, from the \(-5_{\mathrm{th}}\) to the \(+5_{\mathrm{th}}\), as shown in Fig.\ref{fig:fig2}(c). Next, we examine the redistribution of diffracted power across optical diffraction orders at the near-field diffraction monitor. After optimization, the output power distribution shows a strong concentration in the zeroth-order plane wave mode. This indicates that the designed metasurface effectively suppresses higher-order diffraction, guiding complex incident waves into the zeroth-order (normal) direction on the transmission side. 
In contrast, Fig.\ref{fig:fig2}(d) evaluates the performance of the same optimized metasurface when the incident power follows a Gaussian distribution across the diffraction orders. Despite this power distribution, the optimized metasurface consistently directs a significant portion of the transmitted power into the zeroth-order diffraction mode, demonstrating its adaptability and effectiveness in mitigating higher-order optical diffraction. Furthermore, Fig.~\ref{fig:fig2}(e) shows the performance with an inverted Gaussian distribution of incident power. Even though the zeroth-order intensity is zero in the incident wave, the optimized metasurface successfully redistributes higher-order diffraction to the zeroth-order mode. The consistent concentration of power in the zeroth-order diffraction mode under various input conditions demonstrates the robustness of the optimized design and its capability to manipulate higher-order diffraction modes, effectively converting them into a zeroth-order plane wave.

The optimized metasurface offers a versatile solution for advanced optical applications where precise wavefront control is essential. By effectively suppressing higher-order diffraction, it improves light transmission through structured media, making it particularly suitable for UDCs and transparent display technologies. Beyond imaging, its ability to regulate diffraction orders is critical for holography, augmented reality~\citep{yin2021virtual}, and optical computing~\citep{hu2024diffractive}. Additionally, the capability to direct diffracted power into a well-defined propagation mode enhances optical communication~\citep{boffi2004diffraction}, beamforming~\citep{popov2019beamforming}, and integrated photonic circuits~\cite{} by reducing diffraction-induced signal loss.
 
Among these applications, UDC systems present a particularly challenging scenario due to the strong diffraction effects caused by their periodic metal structures. In the following section, we examine how the optimized metasurface can be integrated into UDC architectures to mitigate these effects, enabling improved wavefront restoration and imaging performance.

\section{Wavefront restoration and imaging deblurring in under-display camera systems}

\begin{figure}
    \centering
    \includegraphics[width=1.0\linewidth]{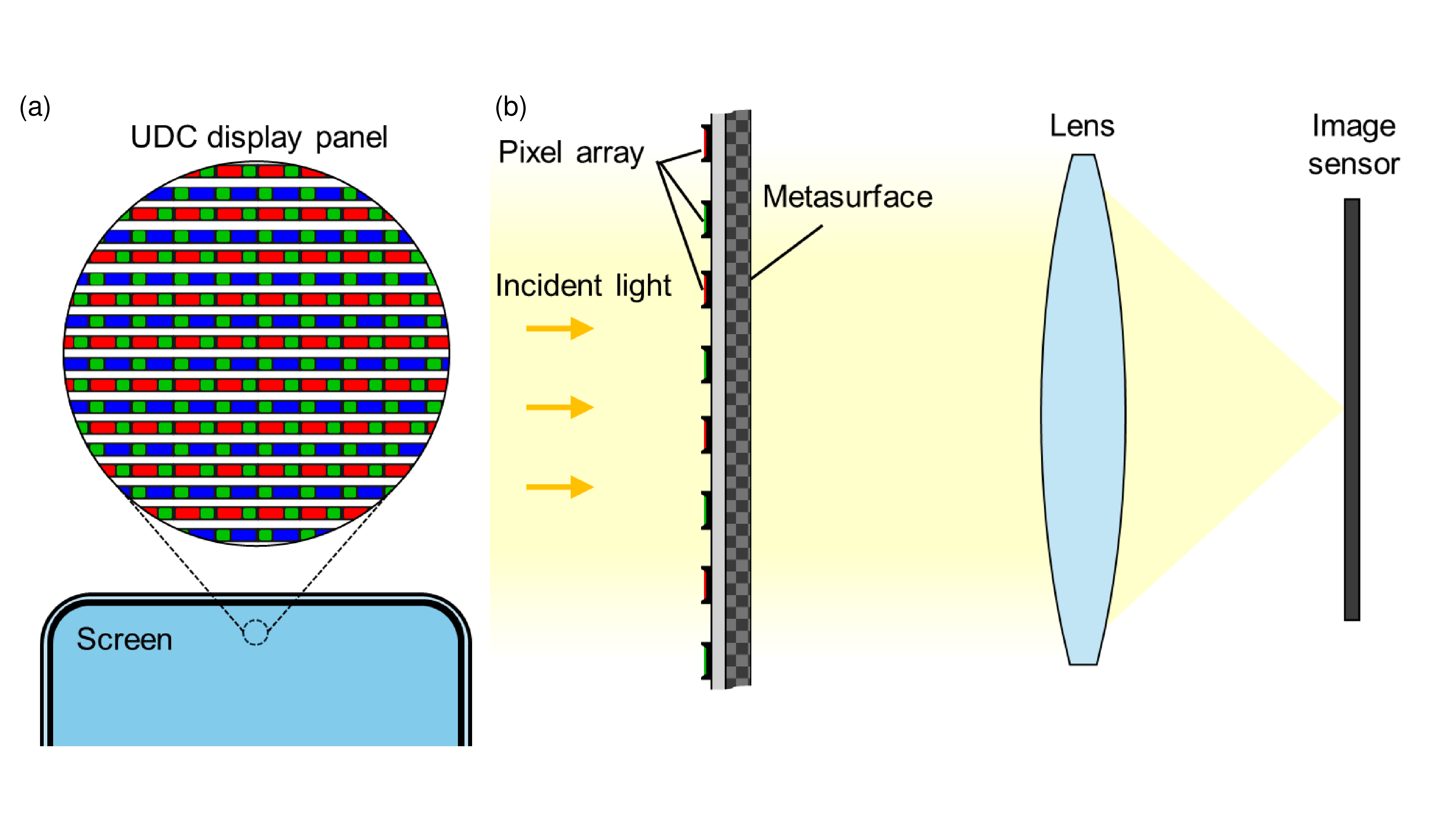}
    \caption{Structural overview of a UDC system and its integration with a metasurface for wavefront restoration. (a) Schematic of a pixelated display panel, where the periodic arrangement of subpixels forms a partially transparent region above the camera module, enabling simultaneous display output and light transmission. However, this periodic pixel structure inherently induces optical diffraction, redistributing incident light into higher diffraction orders and degrading image quality. (b) The UDC system incorporates a metasurface positioned beneath the pixel array to mitigate optical diffraction effects and improve light transmission toward the image sensor.}
    \label{fig:fig3} 
\end{figure}

UDC systems encounter fundamental optical challenges due to the periodic structure of the pixelated display panel. As shown in Fig.~\ref{fig:fig3}(a), the periodic apertures of the metallic pixels within the display panels create partially transparent regions, allowing some of the incident light to pass through while the rest is reflected or diffracted. This periodic aperture structure inherently induces strong optical diffraction effects, which randomly redistribute the incident light into multiple diffraction orders. The presence of uncontrolled higher-order diffraction modes leads to spatial frequency losses, chromatic aberrations, and reduced optical efficiency, all of which significantly degrade real-time imaging performance.

To address these limitations, we propose employing a multi-layer metasurface~\citep{zhou2018multilayer,christiansen2020fullwave,naseri2021generative}, optimized through adjoint optimization in UDC systems. The metallic slab of the metasurface is arranged with a specific periodicity, as shown in Fig.~\ref{fig:fig3}(b). Positioned beneath the pixelated display panel, the metasurface restores the zeroth-order wavefront by suppressing higher-order diffraction modes and directing the transmitted light toward the lens for efficient focusing onto the image sensor. To ensure robust wavefront restoration across multiple visible wavelengths, the metasurface is designed to provide broadband diffraction control. This approach eliminates the strong pattern artifacts typically present in the UDC system without the need for external image restoration algorithms, enabling metasurface-based wavefront correction.

\begin{figure}
    \centering
    \includegraphics[width=1.0\linewidth]{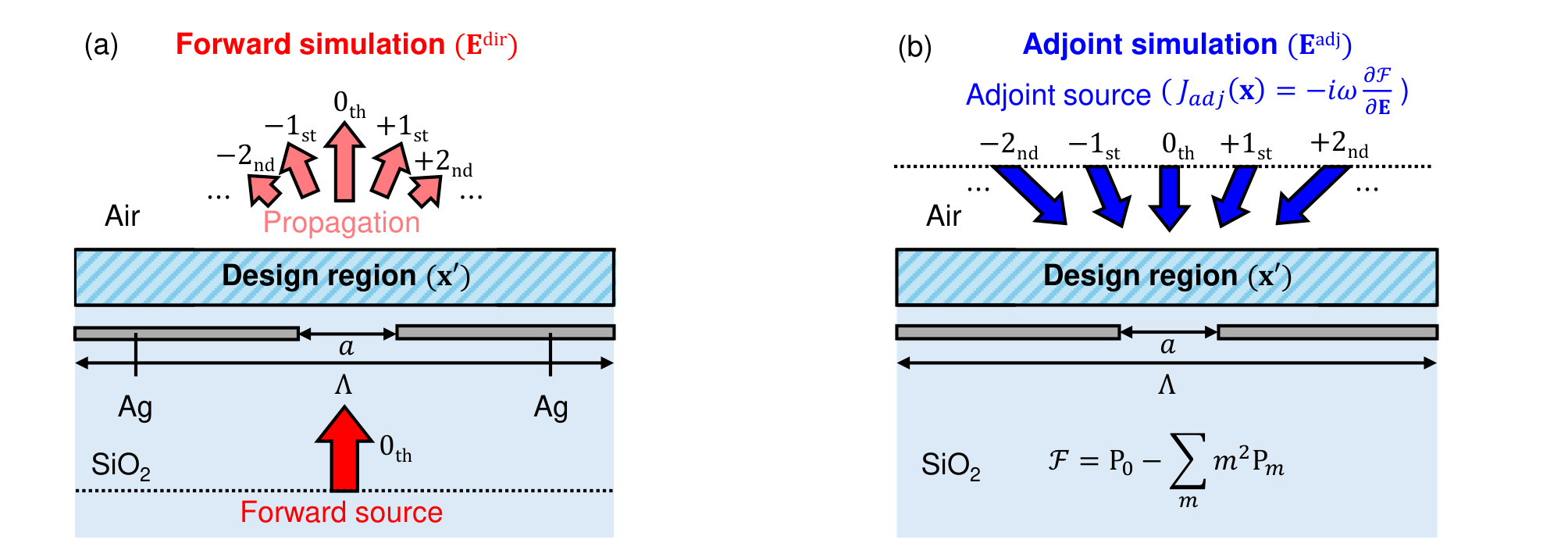}
    \caption{Simulation framework for optimizing metasurfaces under realistic UDC environment. (a) Forward simulation setup, incorporating periodic silver slabs to emulate the metallic cavities of UDC display pixels. The forward source consists of a zeroth-order plane wave, with diffraction behavior naturally arising from the periodicity of the metal slabs. (b) Adjoint simulation setup, where the adjoint source is employed to compute the sensitivity of the figure of merit with respect to design parameters. The metasurface is optimized using a minimax approach to ensure consistent wavefront restoration across multiple visible wavelengths (450~\unit{\nano\metre}, 525~\unit{\nano\metre}, and 630~\unit{\nano\metre}).}
    \label{fig:fig4}
\end{figure}

\subsection{Metasurface integration framework for under-display camera systems}
To optimize the metasurface under realistic UDC conditions, we develop a simulation framework that accurately captures the essential optical behavior of the display panel. As shown in Fig.~\ref{fig:fig4}, the periodic arrangement of silver (Ag) slabs, which mimic the metallic cavities of the UDC display pixels, is incorporated atop the metasurface. Unlike the setup in Section 2, where the input wave included multiple diffraction orders to account for the effects of the periodic structure, this framework uses a forward source consisting solely of a zeroth-order plane wave. In this case, the diffraction behavior is not predefined but instead arises naturally due to the periodic metallic slabs, leading to an uncontrolled redistribution of optical power across multiple diffraction orders.

The metasurface is optimized using an adjoint optimization framework, extending the methodology discussed in the previous section. In this section, we perform two distinct optimization procedures under realistic UDC conditions: a single-wavelength optimization at 940\unit{\nano\metre}, using the same FoM introduced in Section ~\ref{sec:2.2} to assess performance in near-infrared scenarios; and a multi-wavelength optimization across the visible spectrum (450\unit{\nano\metre}, 525\unit{\nano\metre}, and 630\unit{\nano\metre}). Unlike single-wavelength optimization, which may only guarantee performance at a specific frequency, our design adopts a minimax optimization framework that maximizes the worst-case deviation in diffraction control across all three wavelengths. This strategy ensures uniform performance over the full visible range, which is essential for practical implementation in full-color UDC displays.
The FoM for this multi-wavelength optimization is formulated in its epigraph form as~\citep{boyd2004convex}:

\begin{equation}\label{broadband_formulation}
\begin{aligned}
\max \quad &t\\
\text{s.t}. \quad &t \leq \mathcal{F}(\lambda), 
\quad \forall \lambda \in \{450\unit{\nano\metre}, 525\unit{\nano\metre}, 630\unit{\nano\metre}\},
\end{aligned}
\end{equation}
where \(t\) represents an auxiliary variable that captures the worst-case diffraction loss across the three target wavelengths. The wavelength-specific FoM is defined as:
\begin{equation}\label{broadband_FOM}
    \mathcal{F}(\lambda) = P_{0z}(\lambda)-\sum_{m \neq 0} m^2 P_m(\lambda),
\end{equation}
where \( P_{mz}(\lambda) \) represents the power in the \( m_\text{th} \) diffraction mode with \(z\)-polarization electric field at a wavelength \(\lambda\). The multi-wavelength optimization framework then seeks to maximize the worst-performing \(\mathcal{F}(\lambda)\) across the three target wavelengths, ensuring robust diffraction control and wavefront restoration over the visible spectrum. 

\begin{figure}
    \centering
    \includegraphics[width=1.0\linewidth]{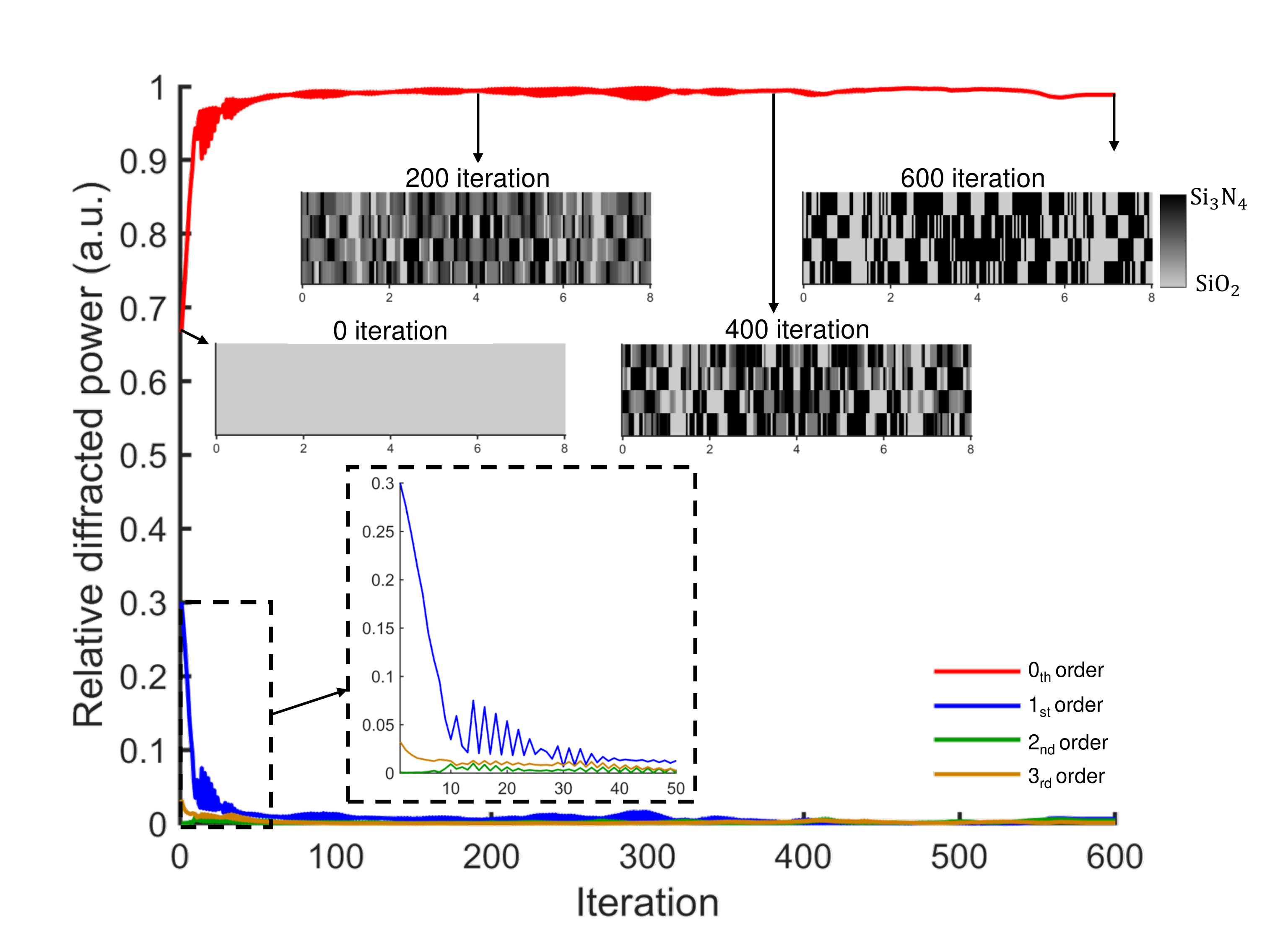}
    \caption{Adjoint optimization iteration versus relative diffraction power at a 940~\unit{\nano\metre} wavelength. The grayscale subfigures illustrate the metasurface design at each iteration, where white-gray represents $\mathrm{SiO_2}$ and black corresponds to $\mathrm{Si_3N_4}$. The zeroth-order diffraction mode increases and stabilizes above 95\% around the 50th iteration, indicating successful wavefront restoration. Simultaneously, higher-order diffraction powers are effectively suppressed. From the 50th iteration to the final iteration, the penalization method is employed to convert the grayscale material permittivity distribution into a binary scale of $\mathrm{Si_3N_4}$ and $\mathrm{SiO_2}$, ensuring manufacturability and optimal performance.}
    \label{fig:fig5}
\end{figure}

\subsection{Performance Analysis of Wavefront Restoration and Imaging Enhancement in UDC Systems}
To evaluate the effectiveness of the proposed metasurface design under practical UDC scenarios, we assess both the single-wavelength and multi-wavelength optimization results through wavefront distribution analysis, diffraction order decomposition, and imaging quality metrics. Section~\ref{sec:single_wavelength_result} focuses on performance at 940~\unit{\nano\metre}, which is relevant for near-infrared applications such as biometric sensing~\cite{}. We visualize the electric field profile and quantify diffraction suppression across orders with and without the metasurface.
Section~\ref{sec:multi_wavelength_result} extends the analysis to visible wavelengths to ensure broadband diffraction control. We evaluate the diffraction power distribution and image quality based on the computed MTF. Comparisons are made between systems with and without the metasurface to demonstrate its effectiveness in practical UDC implementations.

\begin{figure}[!t]
    \centering
    \includegraphics[width=1.0\linewidth]{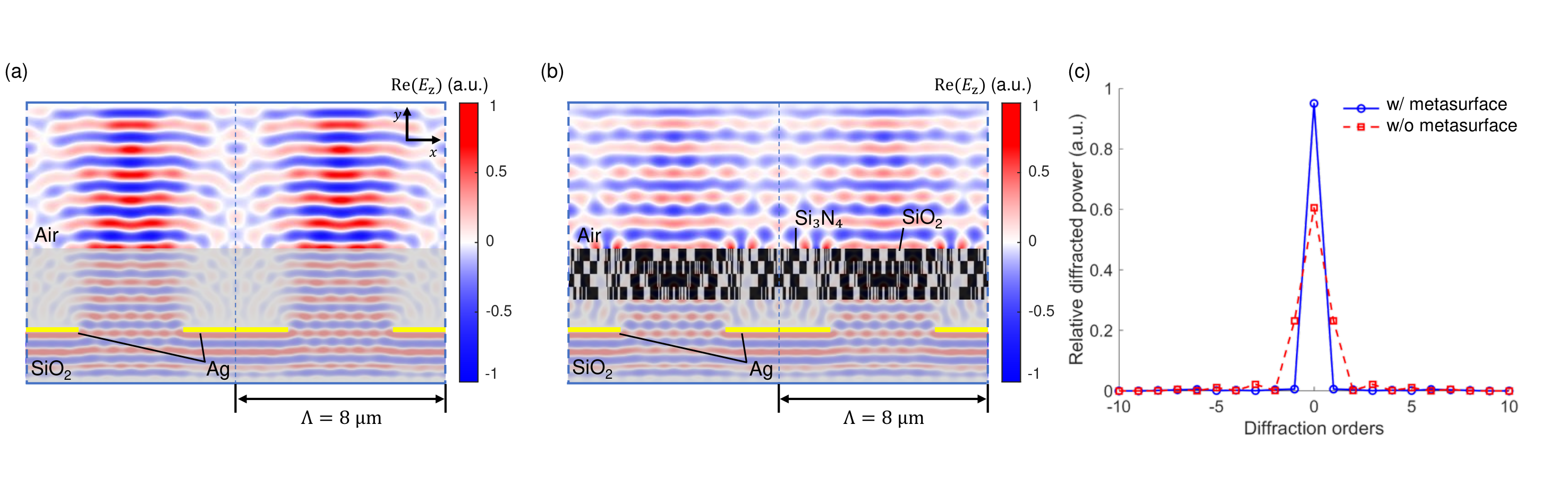}
    \caption{Electric field and diffraction power distribution in a metasurface-integrated UDC system at 940~\unit{\nano\metre}. (a) In the absence of the metasurface, the periodic Ag slabs induce strong optical diffraction, resulting in complex and spatially nonuniform field distributions. (b) With the application of the optimized metasurface, the field becomes collimated and spatially coherent, owing to the suppression of higher-order diffraction. (c) The corresponding diffraction power distribution confirms the efficient redistribution of energy into the zeroth-order mode, thereby validating the metasurface's capability to restore the wavefront.}
    \label{fig:fig6}
\end{figure}

\subsubsection{Wavefront restoration at 940~\unit{\nano\metre} wavelength}\label{sec:single_wavelength_result}
We begin by analyzing the performance of the optimized metasurface under realistic UDC conditions at a single wavelength of 940~\unit{\nano\metre}, which is relevant for near-infrared applications. The evaluation focuses on wavefront restoration efficiency and suppression of higher-order diffraction modes.

Figure~\ref{fig:fig5} presents the evolution of the relative power in diffraction orders over 600 optimization iterations. The zeroth-order diffraction power (red curve) increases sharply during the initial iterations and gradually saturates above 95\%, indicating successful wavefront restoration. In contrast, the higher-order components (blue, green, and orange) rapidly diminish in amplitude, demonstrating effective suppression of unwanted diffraction.

Structural snapshots at selected iterations (0, 200, 400, and 600) illustrate the progressive convergence of the metasurface geometry. The initial uniform distribution evolves into a well-defined binary pattern composed of alternating Si\textsubscript{3}N\textsubscript{4} and SiO\textsubscript{2} regions. This pattern enables precise phase control to redirect incident energy into the desired transmission mode.

 Figure~\ref{fig:fig6}(a) illustrates the spatial profile of the $E_z$ component of the electric field in the absence of the metasurface. The incident plane wave, when passing through the periodic Ag slabs, undergoes strong diffraction, resulting in multiple angularly dispersed wave components. These diffraction-induced components interfere constructively and destructively in space, forming irregular and nonuniform field patterns in the transmission region. The lack of a dominant propagation direction and the presence of higher-order diffraction lobes indicate significant wavefront distortion.
 
In contrast, as shown in Fig.~\ref{fig:fig6}(b), the integration of the optimized metasurface above the Ag slabs significantly alters the field distribution. The metasurface reshapes the diffracted wavefront by imposing a spatially varying phase profile, effectively canceling the phase mismatch introduced by the periodic apertures. As a result, the transmitted field becomes highly collimated and directionally concentrated along the normal (zeroth-order) axis. The restored wavefront exhibits enhanced spatial coherence and planar phase fronts, indicating successful suppression of higher-order diffraction modes and preservation of the input spatial information.

The corresponding diffraction power spectrum is shown in Fig.~\ref{fig6}(c). Without the metasurface, a significant portion of the incident energy is diffracted into higher diffraction orders, reducing overall transmission efficiency. With the metasurface, the power is tightly focused into the zeroth-order mode, with negligible contributions from other orders. This result validates the ability of the metasurface to restore the original wavefront by mitigating the diffraction induced by the structure of the display pixel.

These results demonstrate that the optimized metasurface operates effectively even at 940~\unit{\nano\metre}, providing robust diffraction control and paving the way for its application in NIR-based UDC scenarios, such as biometric sensing and low-light imaging.

\begin{figure}[!ht]
    \centering
    \includegraphics[width=1.0\linewidth]{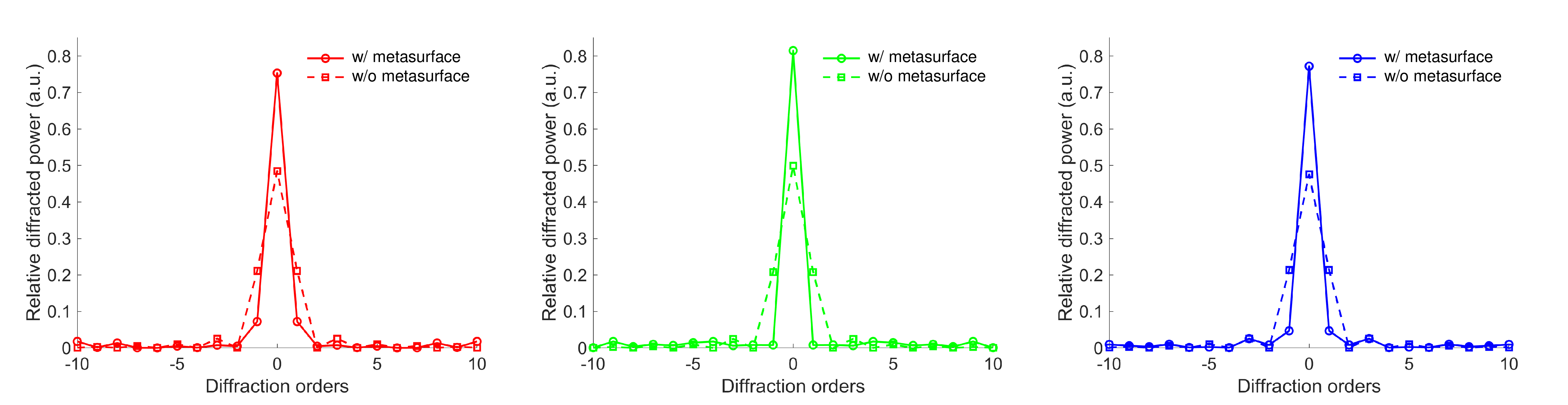}
    \caption{Redistribution of diffracted waves using a metasurface via minimax-based adjoint optimization for wavelengths of 450~\unit{\nano\metre}, 525~\unit{\nano\metre}, and 630~\unit{\nano\metre}. In the absence of the metasurface, significant higher-order diffraction effects are observed across all wavelengths, which may lead to undesirable pattern artifacts in imaging. After applying the optimized metasurface, the majority of incident power is efficiently redirected into the zeroth-order mode at each wavelength. These results demonstrate the multi-wavelength functionality of the metasurface within the visible spectrum and its effectiveness in mitigating wavelength-dependent diffraction effects, thus enhancing the performance of practical UDC applications.}
    \label{fig:fig7}
\end{figure}

\begin{figure}[H]
    \centering
    \includegraphics[width=1.0\linewidth]{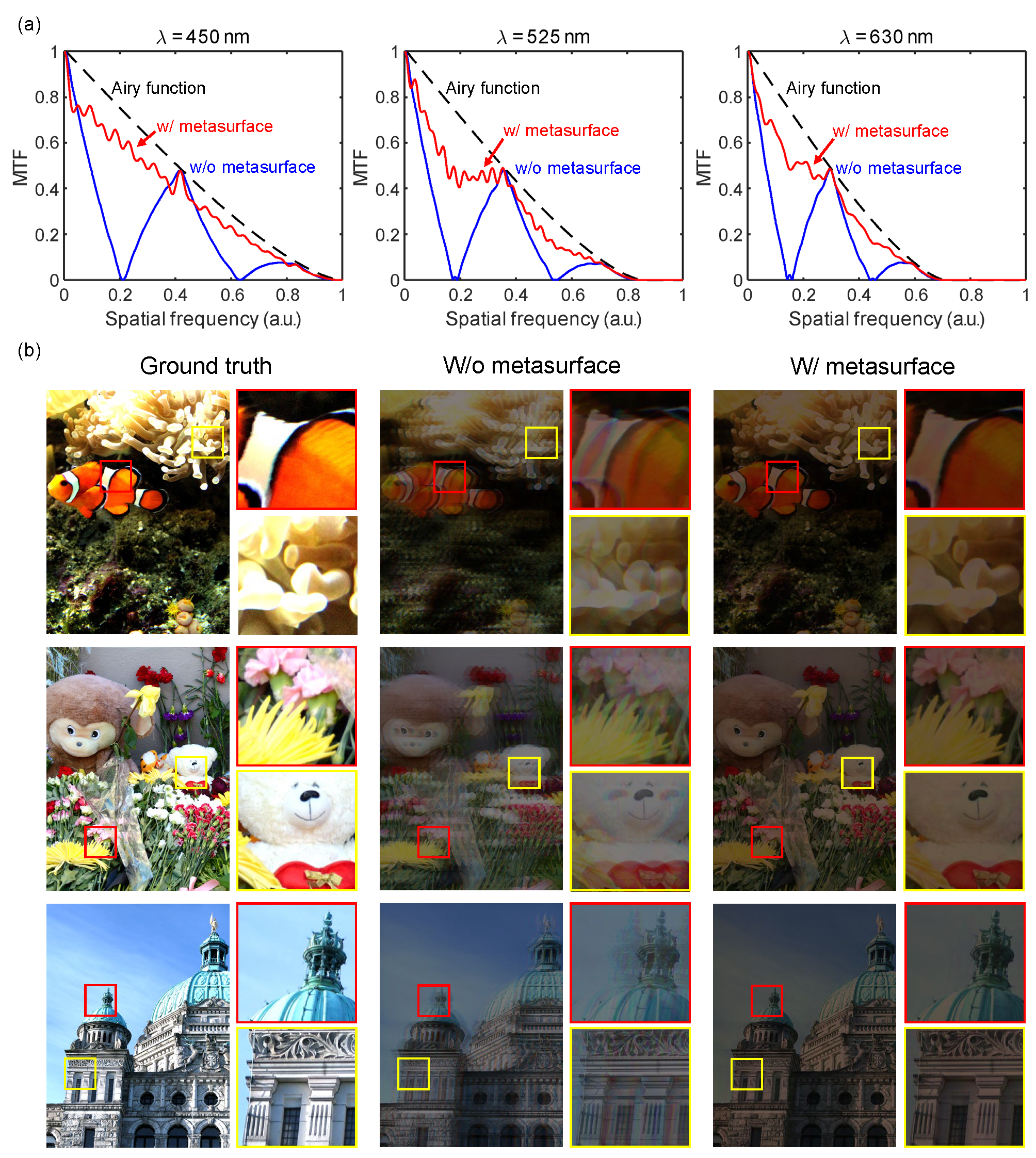}
\caption{Performance evaluation of a metasurface-integrated UDC systems. 
    (a) MTF comparison at three wavelengths (\(\lambda = 450\) nm, \(525\) nm, and \(630\) nm), demonstrating improved spatial frequency response with the metasurface (red) compared to the system without a metasurface (blue). The theoretical Airy function is shown for reference (black dashed line). 
    (b)Visual reconstruction results for various test images. The first column shows the ground truth images, while the second and third columns present the reconstructed images without and with the metasurface, respectively. The inclusion of the metasurface significantly reduces diffraction-induced artifacts and enhances image clarity, as highlighted in the zoomed-in regions (yellow and red boxes).}
    \label{fig:fig8}
\end{figure}

\subsubsection{Multiple wavelengths diffraction correction and imaging deblurring}\label{sec:multi_wavelength_result}
To assess the broadband performance of the proposed metasurface, we conduct a comprehensive multi-wavelength analysis at 450~\unit{\nano\metre}, 525~\unit{\nano\metre}, and 630~\unit{\nano\metre}. This evaluation includes diffraction power redistribution and quantitative imaging performance metrics such as the MTF and PSF.

Figure~\ref{fig:fig7} presents the diffraction power distribution for each wavelength, comparing systems with and without the metasurface. In the absence of the metasurface, strong diffraction induced by the periodic Ag pixel structure leads to significant energy spread across multiple diffraction orders. For all three wavelengths, power is distributed into first- and higher-order modes, resulting in substantial deviation from ideal zeroth-order transmission. In contrast, the metasurface designed through minimax-based optimization robustly concentrates most incident power into the zeroth-order mode across the entire RGB spectrum. Higher-order components are suppressed to near-zero levels, and zeroth-order peaks appear sharper and more dominant. These results indicate that the optimized phase distribution compensates for wavelength-dependent diffraction characteristics, ensuring broadband diffraction control under realistic UDC conditions.

To further evaluate the effect of this broadband diffraction control on imaging performance, we analyze the PSF and MTF of the system, which quantify the spatial resolution and contrast transfer characteristics of an optical system. We analyze Fig.\ref{fig:fig8}, which presents the MTF curve and MTF-based simulated imaging results for the UDC system with and without the metasurface. These evaluations provide a comprehensive assessment of the metasurface’s ability to restore the zeroth-order wavefront and improve spatial frequency transmission, all of which contribute to enhanced imaging performance.

The PSF and MTF serve as fundamental tools in evaluating the optical performance of an imaging system. The PSF characterizes the response of an optical system to an ideal point source, capturing how diffraction, aberrations, and system imperfections affect image formation. Ideally, a point source should produce a sharply localized intensity distribution in the image plane. However, wavefront distortions and diffraction effects cause the energy to spread, resulting in image blur and reduced spatial resolution. Mathematically, the PSF is defined in spatial coordinates and serves as the impulse response of the optical system. A broader PSF indicates greater spreading of the point source, corresponding to degraded resolution, whereas a narrower PSF implies improved optical performance. The shape and extent of the PSF are influenced by diffraction limits, optical aberrations, and wavefront control. In a diffraction-limited system, the PSF takes the form of an Airy disk, while systems with aberrations or diffraction-induced distortions often exhibit broader or asymmetric intensity profiles.

The MTF quantifies the ability of the optical system to preserve spatial frequencies, providing a representation of the retention of image contrast in the frequency domain. As the Fourier transform of the PSF, the MTF describes how well different spatial frequency components are preserved, offering a direct measure of resolution and contrast degradation due to diffraction. A higher MTF at a given frequency indicates better contrast preservation and finer detail retention, while a rapid decline in MTF suggests a loss of high-frequency components, resulting in blurred image and reduced sharpness.

To further quantify the impact of the metasurface on spatial frequency preservation, Fig.~\ref{fig:fig8}(a) presents the MTF curves for three representative wavelengths (450 nm, 525 nm, and 630 nm) in UDC systems with and without the metasurface. The black dashed line represents the Airy function, indicating the theoretical diffraction-limited performance of an ideal imaging system. Without the metasurface (blue curve), the MTF exhibits a steep decline at lower spatial frequencies, indicating a rapid loss of contrast due to the diffraction from the periodic pixel structure. This reduction extends to high spatial frequencies, where significant attenuation confirms that fine image details are severely degraded due to uncontrolled diffraction. Conversely, with metasurface integration (red curve), the smoother MTF profile and its closer alignment with the Airy function indicate that metasurface effectively suppresses diffraction-induced distortions and restores the wavefront, leading to enhanced resolution. Across multiple wavelengths, the metasurface consistently improves contrast preservation, demonstrating broadband diffraction control and wavefront restoration.

Figure~\ref{fig:fig8} (b) presents the MTF-based simulated image results, comparing the optical performance of the UDC system with and without the metasurface. These images were generated by applying the calculated MTFs as convolution kernels to the original high-resolution input images, simulating the effects of diffraction and optical aberrations on image formation. The first column represents ground-truth images that serve as ideal references without optical distortions. The second column shows images generated using the MTF of the system without the metasurface, where diffraction from the pixelated display panel significantly degrades image quality. The steep decline of the MTF curves observed in Fig.\ref{fig:fig8}(a) results in severe blurring, reduced contrast, and chromatic aberration.
In contrast, the third column presents images simulated using the MTF of the metasurface-integrated system. The enhanced contrast at higher spatial frequencies leads to notable improvements in image sharpness and detail preservation. The metasurface effectively suppresses diffraction-induced optical aberrations, enhancing spatial resolution and reducing chromatic distortions across the visible spectrum. These improvements demonstrate the ability of the metasurface to enhance optical resolution and improve image quality by mitigating diffraction effects.
The consistency between MTF enhancements in Fig.~\ref{fig:fig8}(a) and the observed improvements in Fig.~\ref{fig:fig8}(b) confirms that the metasurface successfully restores spatial frequency transmission, suppresses unwanted diffraction orders, and improves imaging performance in UDC applications.

\section{Conclusions}
In this study, we have demonstrated the potential of inverse-designed metasurfaces to address the critical optical challenges in UDC systems, particularly the unwanted optical diffractions caused by the periodic metallic pixel arrays in display panels. By leveraging the adjoint optimization method, we have developed metasurfaces capable of effectively suppressing higher-order diffraction modes and restoring the zeroth-order wavefront, thereby enhancing imaging quality without the need for computationally intensive image restoration algorithms.

Our results show that the optimized metasurface significantly improves wavefront control and imaging performance across multiple visible wavelengths, as evidenced by the enhanced MTF and PSF metrics. The metasurface not only mitigates diffraction-induced artifacts but also preserves spatial frequency information, resulting in sharper and more detailed images. Furthermore, the integration of metasurfaces into UDC systems provides a scalable and manufacturable approach to overcoming the limitations of conventional display architectures. Their compatibility with nanofabrication techniques ensures practical deployment in consumer electronics, enabling full-screen designs with minimal visual disruption. Beyond UDC applications, the demonstrated wavefront control capabilities of metasurfaces present a promising solution for a wide range of optical applications, including holography, augmented reality, and optical computing.

Future research could explore the extension of this approach to broader wavelength ranges and more complex optical systems, further expanding the applicability of metasurfaces in next-generation technologies.

\section*{Declaration of Competing Interest}
The authors declare that they have no known competing financial interests or personal relationships that could have appeared to influence the work reported in this paper.

\section*{Acknowledgements}
This work was supported by the National Research Foundation of Korea (NRF) grant funded by the Korean government (MSIT) under the following grant numbers: (RS-2024-00338048), and (RS-2024-00414119). It was also supported by the Global Research Support Program in the Digital Field (RS-2024-00412644) under the supervision of the Institute of Information and Communications Technology Planning \& Evaluation (IITP), and by the Artificial Intelligence Graduate School Program (RS-2020-II201373, Hanyang University), also supervised by the IITP. Additionally, this research was supported by the Artificial Intelligence Semiconductor Support Program (IITP-(2025)-RS-2023-00253914), funded by the IITP, and by the Korea government (MSIT) grant (RS-2023-00261368, RS-2025-02218723, RS-2025-02283217). This work received support from the Culture, Sports, and Tourism R\&D Program through a grant from the Korea Creative Content Agency, funded by the Ministry of Culture, Sports and Tourism (RS-2024-00332210).

\section*{Data availability}
Data will be made available on request.

\bibliography{UDC}

\end{document}